\documentclass[sigconf]{acmart}
\usepackage[english]{babel}
\usepackage[utf8]{inputenc}

\begin{document}
\title{On Analyzing Self-Driving Networks: \\A Systems Thinking Approach}
\author{Touseef Yaqoob}
\affiliation{\institution{Information Technology University Lahore, Pakistan}}
\email{msee17010@itu.edu.pk}
\author{Muhammad Usama}
\affiliation{\institution{Information Technology University Lahore, Pakistan}}
\email{muhammad.usama@itu.edu.pk}
\author{Junaid Qadir}
\affiliation{\institution{Information Technology University Lahore, Pakistan}}
\email{junaid.qadir@itu.edu.pk}
\author{Gareth Tyson}
\affiliation{\institution{Queen Mary University of London, United Kingdom}}
\email{g.tyson@qmul.ac.uk}
\renewcommand{\shortauthors}{Yaqoob et al.}

\begin{abstract}
\sloppy

The networking field has recently started to incorporate artificial intelligence (AI), machine learning (ML), big data analytics combined with advances in networking (such as software-defined networks, network functions virtualization, and programmable data planes) in a bid to construct highly optimized self-driving and self-organizing networks. It is worth remembering that the modern Internet that interconnects millions of networks is a `complex adaptive social system', in which interventions not only cause effects but the effects have further knock-on effects (not all of which are desirable or anticipated). We believe that self-driving networks will likely raise new unanticipated challenges (particularly in the human-facing domains of ethics, privacy, and security). In this paper, we propose the use of insights and tools from the field of ``systems thinking''---a rich discipline developing for more than half a century, which encompasses qualitative and quantitative nonlinear models of complex social systems---and highlight their relevance for studying the long-term effects of network architectural interventions, particularly for self-driving networks. We show that these tools complement existing simulation and modeling tools and provide new insights and capabilities. To the best of our knowledge, this is the first study that has considered the relevance of formal systems thinking tools for the analysis of self-driving networks. 


\end{abstract}


\maketitle

\section{Introduction}
\sloppy
\label{sec:introduction}

The exponential growth in the number of connected devices and users in networks is placing significant stress on current human-in-the-loop network management architectures. There is now interest in equipping networks with autonomous run-time decision-making capability through the incorporation of machine learning (ML), big data network analytics, and network telemetry to allows networks to configure, manage, and heal themselves. The idea that networks should learn to drive themselves is gaining traction, and there is a lot in the networking community to develop \textit{self-driving networks} \cite{feamster2017and}. 

The idea itself is not entirely new and reflects a recurring motif seen in various guises such as cognitive networking \cite{thomas2006cognitive}, self-organized networks \cite{aliu2013survey}, knowledge defined networks \cite{mestres2017knowledge}, and most recently, data-driven networking \cite{jiang2017unleashing} and \textit{self-driving networks} \cite{feamster2017and,kompella2017}. The vision of self-driving networks is promising, and finds much encouragement from recent advances in ML (such as deep learning) and networking (such as software-defined networks, programmable data planes, and edge computing). The real concern is not to only see the potential benefits of this approach, but to also critically understand potential downsides---in this work we seek to investigate the pros and cons of self-driving networks using \textit{systems thinking}.

It's worth noting that modern networks, and their integration into the global Internet, yields a \textit{complex adaptive social system} that encompasses the interaction of a vast diversity of autonomous devices, human users, applications, and service providers. Due to the presence of intertwined nonlinear feedback loops, complex systems often work counterintuitively with effects and causes linked indirectly through circuitous paths, distant in time and space, which makes the job of analyzing the efficacy of an intervention difficult and error-prone. With the emergence of self-driving networks, these feedback loops will become even more intricate and intertwined, which will mean that it will be harder to nudge the system towards desired behavior except through a deeper understanding of the underlying system. Our mental models or conventional modeling tools are wholly incapable of accurately tracking these locked feedback effects and need to be supplemented by systems thinking tools. 




\subsection{Systems Thinking: Closing the Loop}

\begin{table*}[!hbtp]
\footnotesize
\caption{Comparing Conventional vs. Systems Thinking (Details: \cite{senge2006fifth}\cite{stroh2015systems})}
\label{conventionalVsSystems}
\begin{tabular}{p{3.5cm}|p{6cm}|p{7cm}}
& \textbf{\textit{Conventional Thinking}} & \textit{\textbf{Systems Thinking}} \\\hline

\textit{Model of thinking} & Linear, causal, open-looped, immediate feedback & Nonlinear, multi-causal, close-looped with delayed feedback \\\hline
\textit{Determining a problem's cause} & Obvious and easy to trace & Indirect and non-obvious \\\hline
\textit{Cause of problems} & External to the system & Internal (System-as-a-cause thinking) \\\hline
\textit{How to optimize?} & By optimizing the parts & By optimizing relationships among the parts \\\hline
\textit{Where to intervene?} & Aggressive use of ``obvious'' solutions & Careful change applied at the ``leverage points'' \\\hline
\textit{How to resolve problems?} & Cure the symptoms & Fix the systemic causes \\\hline

\end{tabular}
\end{table*}

We can define a system as, ``\textit{an interconnected set of elements that is coherently organized in a way that achieves something}''---a definition given by Donella Meadows, a highly influential system thinker and the lead author of the best-selling ``Limits to Growth'' \cite{meadows2004limits}. This ``something'' may not be what the designers had in mind. There's a belief in systems thinking that systems are perfectly designed to achieve the results they are currently achieving. Systems thinking may be defined as ``\textit{the ability to understand the systemic interconnections in such a way as to achieve a desired purpose}'' \cite{stroh2015systems}. Thus systems thinking can be used to see more clearly the purpose the system is accomplishing and to reflect on its deviation from the defined purpose and to effectively reconcile. 

The following four key distinctive thinking patterns distinguish systems thinking from conventional thinking \cite{richardson2011reflections}: \textit{firstly}, the ability to think dynamically (e.g., using graphs over time); \textit{secondly}, to think causally through feedback loops; \textit{thirdly}, to think of stocks and flows (i.e., accumulation and transfer); and \textit{finally}, to think more deeply about endogenous influences (where the system itself is the cause of the observed problems). System thinking can also be understood by contrasting it with open-loop based conventional thinking (see Table \ref{conventionalVsSystems}), which fails to take into account that social systems are more properly modeled as \textit{multi-loop nonlinear} feedback systems and in such systems hardly anything is ever influenced linearly in just one direction \cite{forrester1971counterintuitive} and in nonlinear systems, ``\textit{the act of playing the game has a way of changing the rules}'' \cite{gleick2011chaos}.

\subsection{Systems Thinking For Self-Driving Networks: Motivation and Importance}


\subsubsection{Facilitating ``system-as-cause'' thinking} 

To paraphrase management expert, one cannot be part of the solution, if one is unaware of how one is part of the problem. In systems thinking, it is considered an axiom that every influence is both a cause and an effect---i.e., it is possible that when A causes B, B also causes A through a feedback loop. In other words, in such doubly looped system, systems cause their own behavior endogenously. We can use this \textit{system-as-a-cause} understanding in self-driving networks along with system thinking tools to anticipate how our system goals may be causing unanticipated problems and then work towards ensuring that the purposes achieved by the self-driving network are  congruous to our stated goals.

\subsubsection{Support for rigorous big picture thinking} It affords us an ability to see the big picture by expanding our \textit{time} and \textit{thought} horizons. Using system thinking tools, we can take better policy decisions regarding self-driving networks and avoid an exclusive reliance on implicit mental models, which are ill-suited for this task since they are simplistic (since they inadvertently substitute a higher-order nonlinear system for a linear causal one); \textit{narrow} (i.e., not broad enough to see the big picture); and \textit{myopic} (since they tend to discount the future and focus predominantly on the short-term) \cite{forrester1971counterintuitive}. 

\subsubsection{Identification of leverage points and the system structure} It allows us to see the leverage points manipulating which we can produce better results with fewer resources in more lasting ways. Systems thinking can also be used to better understand the connections between the various subsystems. In particular, it helps us identify non-obvious connections between effects and causes; and also find missing connections, which if they had existed, would have improved the system performance of our self-driving networks.

\subsubsection{Management of unintended consequences} 
Using systems thinking can help us anticipate and avoid the negative consequences of well-intentioned solutions. This can be done both \textit{prospectively} by anticipating unintended consequences during strategic planning or \textit{retrospectively} by understanding more deeply the non-obvious causes of existing chronic complex social problems.

\subsubsection{Leveraging a rich set of versatile tools} We can leverage tools from a vast library developed by the systems thinking community, which has been active since its genesis at MIT in the 1950s \cite{senge2006fifth}, for use in self-driving networks. The field of systems thinking is a highly-developed discipline with many schools of thought (including system dynamics, complexity theory, general systems theory, human system dynamics, etc. \cite{stroh2015systems} \cite{arnold2015definition}) and highly-developed qualitative and quantitative tools \cite{kim1995systems}---e.g., structural thinking tools \cite{senge2006fifth} \cite{stroh2015systems}).
such as causal loop diagrams, graphical functions diagrams; models using stocks and flows, iceberg, and bathtub; and system archetypes (discussed in section \ref{sec:systemArchetypes}). These tools have been successfully used to study policy-making in other domains such as healthcare, education, management \cite{sterman2000systems} and look promising for self-driving networks as well.

\subsection{Contributions of This Paper} In this paper, we aim to highlight that the Internet and self-driving networks should be envisioned as complex adaptive systems in which we should be wary of easy solutions and quick fixes. As pointed out by H. L. Mencken, there's always an easy solution to every problem that is neat, plausible, \textit{but} wrong. In a similar vein, systems thinking research informs us that most well-intentioned solutions fail to sustainably solve their addressed problems and may actually create more problems than they solve. However, not all solutions are doomed in this manner---some high-leverage solutions exist, which are not constrained by balancing feedback loops and which systems thinking can uncover. We propose the use of tools and insights from systems thinking in self-driving networks for managing the unintended consequences of policies and for devising high-leverage effective solutions. To the best of our knowledge, this is the first proposal to use systems thinking insights and tools for the study of self-driving networks and possibly also for the Internet.

\section{System Laws and Insights}
\sloppy
\label{sec:systemlawsinsights}


\textit{Leverage Points}. Systems do not respond equally to all interventions but rather respond according to the \textit{principle of leverage}. We observe in practice soberingly that relatively few policy interventions in social systems are \textit{high-leverage} (i.e., capable of producing substantial system change) and most interventions are \textit{low-leverage} (i.e., they are not capable of producing a significant change in the long run and will likely also create other problems). It becomes important to seek out these high-leverage intervention points. It turns out that these high-leverage policies points are not where most people expect, and if even identified, they are prone to be altered in the wrong direction by people acting intuitively---thereby highlighting the counterintuitive nature of complex social systems emphasized in system dynamics research \cite{forrester1971counterintuitive}. Donella Meadows, in her essay that ranks places to intervene in a system \cite{meadows2007leverage}, states parameter optimization is typically low-leverage and better results can be obtained by optimizing information flows (e.g., by minimizing information sharing delays) and by changing the rules of the system (i.e., the incentives and the constraints). The most powerful way to change a system, however, Meadows state is to change its goals and mindsets/paradigm, out of which its goals, rules, and culture emerge. We can use these insights about leverage points to unearth the few sensitive influence points in self-driving networks and avoid some of the ruts that traditional networks fell into (explained in section \ref{sec:systemicproblems})
 


\textit{System Laws}. Peter Senge in his book ``The Fifth Discipline'' \cite{senge2006fifth} highlights some laws (recurring themes) observed in the discipline of systems thinking. These laws have been generalized from manifestations in a number of diverse settings, and they serve again as a sombre reminder that systems have a life of their own and they will resist being tampered and will chronically return back to demonstrate the effects that follow from its systemic structure. Some of these laws most pertinent to our work are 1) \textit{today's problems come from yesterday's ``solutions''}; (2) \textit{behavior grows better before it grows worse} (i.e., benefits of quick-fix interventions accrue in the short-time, only to neutralize and worsen off in the long-run); (3) \textit{the easy way out usually leads back in}; (4) \textit{the cure can be worse than the disease} (i.e., short-term improvements can lead to long-term dependencies); (5) \textit{cause and effect are not closely related in time and space}; and (6) \textit{small changes can produce big results---but the areas of highest leverage are often the least obvious}.

\section{Systemic Problems and System Archetypes}
\sloppy
\label{sec:systemicproblems}


A  major insight of systems thinking is that the root causes of chronic complex problems often lay in the underlying systemic structure, which is often non-obvious since causes take an indirect route through nonlinear time-delayed subsystem interactions to create the problematic effect.This means that not all problems are solvable through interventions, some problems are \textit{systemic} and follow from the system's fundamental architectural choices. Systems research also tells us that self-organizing nonlinear feedback systems are inherently unpredictable and not totally controllable---as noted by Neil Postman \cite{neil1992technopoly}, ``\textit{technology is both a burden and a blessing; not either-or, but this-and-that.}''


\subsection{Paradoxes of Internet System Design}

We can use the systems thinking concept of \textit{system-as-a-cause} to explain how the perennial Internet nuisances (such as spam and lack of privacy, security and QoS) are not isolated problems but as noted by Keshav \cite{keshav2018paradoxes} follow endogenously as the byproducts of the Internet's design preferences. This work points out that paradoxically the Internet's architectural elements most responsible for its success are also responsible for its most vexing problems. It is clear that if we want to fix these ancillary problems, this cannot be achieved superficially without changing the systemic causes. This underscores the importance of thinking long-term in the design of self-driving networks and of anticipating such paradoxes outcomes that stem directly from the system design.


\subsection{Tussles, Conflicts, and Dilemmas}

It must be kept in mind that different stakeholders on the Internet ecosystem have different, often conflicting, interests, which when independently pursued create ``tussles'' of various types. Some people wish for privacy on the Internet, others prefer accountability and the ability to identify. Some protocols aim to implement a functionality in an end-to-end manner; others may prefer an in-network mechanism. The functionality implemented at various layers may be neutralized or may even conflict. Thus there is ``\textit{not a single happy family of people''} on the Internet with aligned goals \cite{clark2002tussle}. Apart from tussles and conflicts, Internet protocols and applications also often face dilemmas in which the goals of the subsystem and the overall system conflict. One of the major insights of systems thinking is that the best way to optimize a system is not to independently optimize each subsystem but to optimize the relationships among the parts (which often is the bottleneck). An important implication for self-driving networks is that we cannot be everything to everyone---it becomes important therefore to clearly articulate our goals while keeping in view that different subsystems do not have homogeneous interests or points-of-view. We can also use systems thinking tools to anticipate the non-obvious interactions between the subsystems and use insights therefrom to minimize tussles and bottlenecks.

\subsection{Perils of Unintended Consequences}


Unintended consequences are the staple of complex social systems, which follow unexpectedly from the nonlinear interactions between subsystems \cite{forrester1971counterintuitive} and our propensity to intervene in systems with our ``solutions''---solutions regarding which Eric Sevareid, an economics commentator had astutely noted, ``\textit{the chief source of problems is solutions}'' Our problem-solving instinct also creates a number of followup problems and networking systems (including future self-driving networks) are not immune to this tendency  \cite{raghavan2015abstraction}.

\subsection{System (Misbehavior) Archetypes}
\sloppy
\label{sec:systemArchetypes}

System dynamics literature is rife with examples of fixes gone wrong---in which well-intentioned common-sense interventions to mitigate a particular problem has gone on to aggravate it (not to mention the creation of other problems) \cite{senge2006fifth}. These archetypes provide us valuable information about typical pitfalls that experts have repeatedly noticed. These archetypes are common and easily understood, and once internalized, can help designers and stakeholders in identifying the rut they are in and to identify recognizable paths (the leverage points) taking which will lead to a resolution. All these pitfalls contain lessons that can guide the designers of self-driving networks towards more productive strategies. In particular, paying heed to these archetypes and taking necessary corrective actions will result in behavior that is sustainable and effective (Archetypes \# 1,2,4,5,7), more stable (Archetypes \#6),  and fairer and more equitable (Archetypes \# 3). These system failure archetypes are listed in Table \ref{Fig:SystemArchetypes}, along with some networking examples (due to the shortage of space, these are not always elaborated upon in text). 


\begin{table*}[!hbtp]
\centering
\footnotesize
\caption{System archetypes identified in system dynamics \cite{senge2006fifth} with networking examples}
\label{Fig:SystemArchetypes}
\begin{tabular}{l|l|l|l}
No. & Archetype Name& Description & Networking Examples\\\hline
1  & Fixes That Backfire& A quick solution with unexpected long-term consequences& \cite{kawadia2005cautionary} \cite{keshav2018paradoxes} \cite{gettys2012bufferbloat} \cite{kelly2001mathematical}; IP NAT \\\hline
2  & Limits to Growth& Improvement accelerates and then suddenly stalls& IPv4; \cite{raghavan2011networking} \cite{handley2006internet} \\\hline
3  & Success to the Successful& Things get better for ``winners'' and worse for ``losers''& \cite{benkler2006wealth} \cite{grotker2015citizens}\\\hline
4  & Shifting the Burden& Systems unconsciously favor short-term, addictive solutions& \cite{kawadia2005cautionary} \\\hline
5  & Tragedy of the Commons& Shared unmanaged resource collapses due to overconsumption&
\cite{damsgaard2006wireless} \cite{gupta1997internet}\\\hline
6  & Escalation & Different parties take actions to counter a perceived threat & \cite{nithyanand2016adblocking} \cite{clark2002tussle}\\\hline
7  & Eroding Goals &  Short-term solutions lead to the deterioration of long-term goals& \cite{clark2002tussle} \\\hline

\end{tabular}
\end{table*}

\subsubsection{Fixes that Backfire}

This system archetype is associated with the story of unintended consequences. Fixes that backfire are characterized by the use of a quick fix to reduce a problem symptom that works in the short run but at the cost of long-term consequences, which people often fail to see due to long system delays. Fixes that backfire systems archetype is a common pitfall in networks with some networking examples being  (1) increasing queue buffers to decrease packet loss but instead causing bufferbloat \cite{gettys2012bufferbloat}, and (2) introducing additional links to an existing system only to see overall performance drop (Braess' paradox) \cite{kelly2001mathematical}.


\subsubsection{Shifting the Burden}

This archetype is the system archetype associated with the story of unintended dependence. This system archetype arises from dependence on a quick fix, which is resorted to when the more fundamental solution is too expensive or too difficult to be implemented. This archetype differs from ``fixes that backfire'' since the fundamental solution may not be apparent or applicable in the latter. With the ``shifting the burden`` archetype, the quick fix produces temporary relief by treating the symptoms, which tends to reduce the motivation to implement the more fundamental solution.


\subsubsection{Limits to Growth}

This archetype is the system archetype that describes the story of unanticipated constraints, which underscores the insight that no physical system can sustain growth indefinitely. Any engine of growth, however successful and adroit it may be, will inevitably be constrained by internal and external bottlenecks and constraints---e.g., Meadows showed in the pioneering work \textit{Limits to Growth} \cite{meadows2004limits} that we cannot sustainably support perpetual growth in a finite world. In the field of networking, researchers are now exploring how permanent energy crisis scenario may fundamentally limit our ability to maintain the current-day Internet architecture and what should be our response in such an eventuality \cite{raghavan2011networking} \cite{qadir2016taming}.


\subsubsection{Success to the Successful}

This archetype is associated with the story of the winner taking it all, and it refers to the common tendency in social systems for the privileged to accumulate more of the benefits than the underprivileged. This archetype commonly occurs in system dynamics and helps to make differences in privileges more pronounced over time. For the purpose of self-driving networks, this archetype has implications for policy making for network neutrality and fair usage.


\subsubsection{Eroding Goals}

This archetype, also called \textit{``Drifting goals''}, is another easily recognized system archetype. It is a special case of ``shifting the burden'', where the preferred quick fix is to keep lowering the system goals, and the continuous adjustment to these lowered goals turns out to be fatal for the system. Drifting goals is typically explained through the \textit{metaphor of the boiled frog}---which describes the situation of a frog being dropped into a pot of boiling water from which it immediately pops out to save itself; but if it is put in lukewarm water, which is then gradually heated, the frog keeps adjusting and fails to recognize the danger of the rising warmth thereby killing itself. 


\subsubsection{Escalation}

This system archetype describes the story of unintended proliferation in a sort of an arms race in which the harder you push, the harder the adversary pushes back. An example of this tussle in networking is the consumer profiling by different operators and then selling of this consumer data to advertising agencies \cite{nithyanand2016adblocking}.

\subsubsection{Tragedy of the Commons}

This archetype refers to the story of a depleting collective shared resource that all parties are interested in exploiting but none feel responsible for maintaining.  For networking, this is applicable for unlicensed use of natural shared limited resources such as radio spectrum---e.g., the problem of interference in unlicensed wireless commons \cite{1572368}.

\section{Applying Systems Thinking in Self-Driving Networks}
\sloppy

\begin{figure}[hbtp]
 \centering
 \includegraphics[scale=0.32]{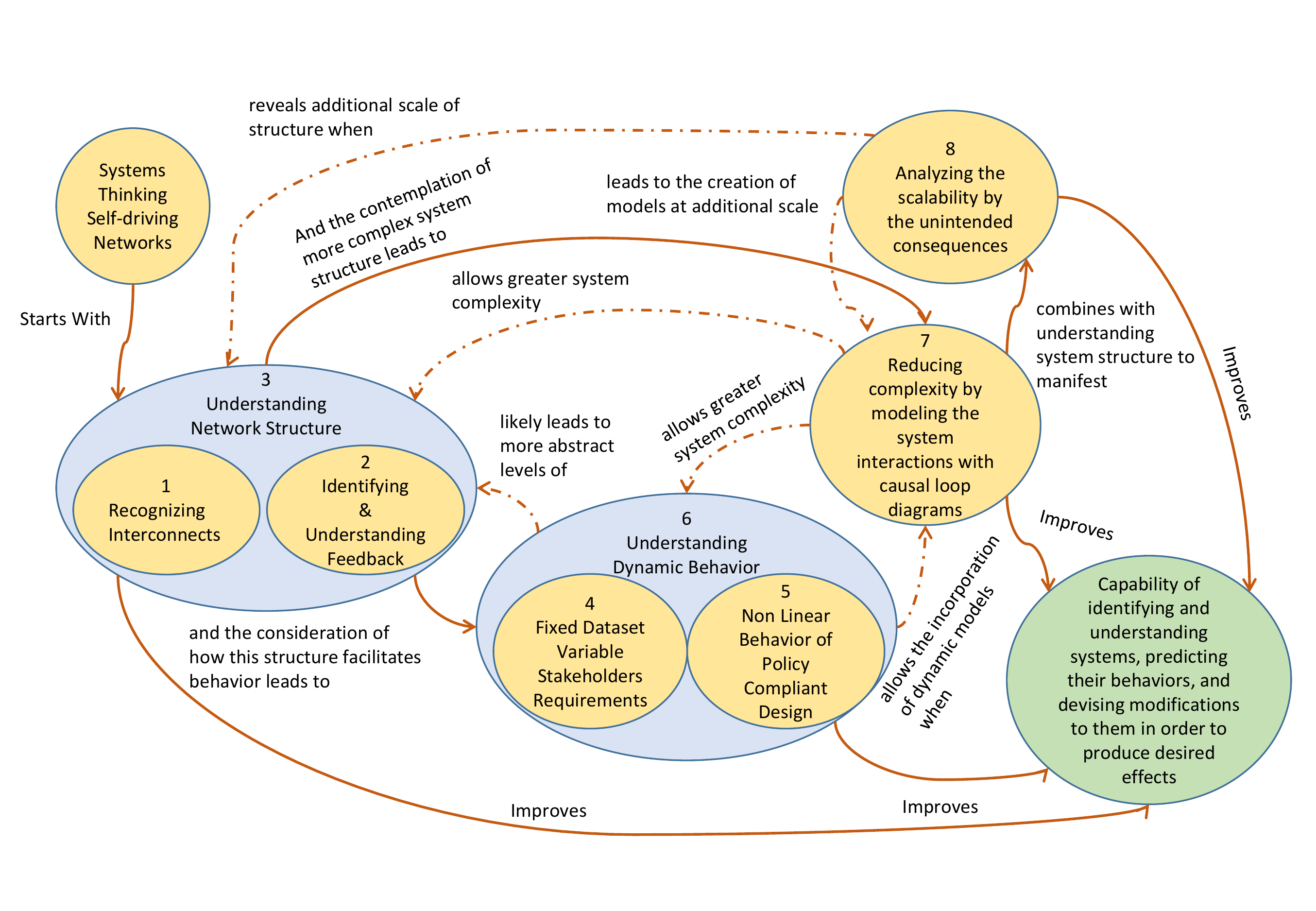}
 \caption{Systemigram of Self-driving Networks}
 \label{fig:systemigram}
\end{figure}

In this section, we begin to explore how systems thinking may be applied to self-driving networks. Our initial foray is exploratory since a detailed description of the various tools system thinking is outside the scope of this short position paper. A systemigram describing how system thinking concepts may be applied in the context of self-driving networks is demonstrated in Figure \ref{fig:systemigram}. 

Our work proposes a new line of thinking and solutions for solving problems in self-driving networks and we recognize that we propose more questions than answers in this paper. We propose that the community should look at systems thinking tools to answer the following fundamental questions relating to self-driving networks:

\begin{enumerate}
    
\item Who are the various stakeholders who will be affected if self-driving networks are introduced? How do they interact and are influenced by others?

\item What are our objectives/goals that we want to achieve from self-driving networks? What ethical and security/privacy challenges can arise as a result of our system goals?

\item What unintended consequences and constraints can we unmask using systemic thinking tools for the different stakeholders?

\item How can concepts of feedback, archetypes, bathtub, iceberg improve our understanding of the systemic structure of self-driving networks?

\item Which interventions and policies in self-driving networks will likely be high leverage?

\end{enumerate}

\subsection{Improving System Structure}
\sloppy



\subsubsection{On Architecting Goals for Networks}

As per Donella Meadows's ranked list of recommended places to intervene in a system \cite{meadows2007leverage},  interventions that aim to optimize parameters are nowhere as powerful or fundamental as interventions that aim at changing the system's goals and paradigms. To ensure better performance, we need clearer articulation of what the goals of our self-driving networks are. 

\subsubsection{Focusing on System Bottlenecks} To paraphrase the words of the prominent system theorist Russell Ackoff, a system's performance is never the sum of the performance of its parts, but the product of their interactions. To improve system performance, bottlenecks should be identified and efforts should be invested in alleviating these bottlenecks rather than on optimizing subsystems separately. In addition to identifying the problematic connections (i.e., bottlenecks), self-driving networks can also leverage systems thinking to determine new connections that can potentially mitigate bottlenecks through more efficient information sharing. 

\subsubsection{Timely sharing of information}
In networks, there can various delays involved that contribute to the feedback loops that exist in a networking system. There may be delays in the transmission of information and the relevant information may not be available at the decision maker/ controller in the case of distributed systems. To facilitate the required timely sharing of information, new architectures and strategies (such as split control architectures) are needed \cite{jiang2017unleashing}.





\subsection{Finding the Right Functional Split}
\sloppy
Despite the moniker of self-driving networks, humans will not be removed completely from the management of networks. There will inevitably remain be a functional split between humans and computers for network management. It is true that algorithms can prevent many of the trivial manual mistakes that can afflict network operations, but it's worth keeping in mind that algorithms are also not impervious to blunders \cite{kugler2016happens} (since algorithms do not have the common sense and can only learn from the given instructions or data). With it being well known that human intuition is sometimes marvelous and sometimes flawed \cite{kahneman2009conditions}, an important (and not entirely technical) exercise is to map the boundary conditions for the management of self-driving networks where we can safely relegate matters and operations to algorithms and where we will like to have human oversight (e.g., in crafting policies related to matters pertaining to ethics and human subjects). There will likely be many configurations of self-driving networks and more debate is needed on the right functional split---especially to avoid reliability, security, and ethics related problems.

\subsection{Ethical Challenges}
\label{sec:challenges}

Giving away the agency of decision-making to algorithms in self-driving networks opens up a plethora of ethical challenges. Despite the many successes of ML, experts have pointed out that many modern ML techniques work like a black-box and may make predictions without really knowing why. The harmful effects of opaque unregulated ML-based algorithms described by O'Neil in \cite{o2017weapons} represent a significant concern for self-driving networks. In \cite{feamster2017and}, an example of ML-based spam filter was proposed that used features such as the autonomous system number of the sender; although very useful, ML algorithms are not perfect and one should reason ahead about the potential of ``false positives'' and take steps to ensure that we do not inadvertently create ``weapons of math destruction'' or strengthen existing stereotypes \cite{o2017weapons} \cite{yapo2018ethical}. Systems thinking can help us perform higher order thinking and determine unintended consequences of relying on opaque ML algorithms and potentially biased datasets. 

The question of \textit{agency}---i.e., ``\textit{who will take the ethical decision?}---also looms large for self-driving networks and it's not clear if network operators and managers should make ethical decisions on behalf of the uses and if so then how. These ethical questions may not have an objectively straightforward resolution and present dilemmas (e.g., self-driving network version of trolley problems \cite{nyholm2016ethics} may arise in which the interest of many might be vying with the actions of a limited few and one has to decide how this conflict is to be addressed). The ethical decisions adopted may also have a strong social and economic implications, as the policy may be beneficial for some stakeholders but not for others, and through the change in incentives may trigger changes in the services and products the clients will use. Systems thinking can allow us to rigorously study these ripple effects in self-driving networks. Ethical concerns related to networking research are now being documented and guiding principles articulated \cite{alllman2007issues, jones2015ethical}, but specific ethical concerns around self-driving networks require more deliberations.

\subsection{Security Challenges}
\label{sec:challenges}


As remarked tellingly by Russell Ackoff, ``\textit{no problem stays solved in a dynamic environment}.'' Since algorithms are trained using historical datasets, self-driving networks are always vulnerable to future evolved adversarial attack. One should try to use systems thinking tools to anticipate the various kinds of crippling attacks that adversarial attackers can launch on self-driving networks. Relying on algorithmic measures also opens up an opportunity for malicious applications/users to game the system. According to the \textit{Campbell's law}, developed by the social scientist Donald Campbell, the more any quantitative social indicator is used for social decision-making, the more it will be subject to corruption pressures and the more likely will be to distort the social processes it is intended to monitor. 

\section{Conclusions}
\sloppy
\label{sec:conclusions}


Our technological interventions in the Internet have wide-ranging implications since Internet technologies are deeply embedded in a larger social, political and cultural context. With the rise of interest in self-driving networks, which will become part of the larger Internet, there is a need to rigorously look at how these technologies will affect---positively as well as negatively---all the stakeholders. In order to devise appropriate policies for future self-driving networks, it is essential that we not only use traditional machine learning (ML) and analytic tools but also complement these with systems thinking tools to study the dynamics of interaction within self-driving networks and between it and other interacting systems. We believe that system thinking complements traditional methods (e.g., mathematical/statistical/ML models as well as discrete-event simulators) to bring unique insights not attainable through these other methodologies. Our work applies for the first time powerful insights from systems thinking and demonstrates their relevance for studying the broad implications of self-driving networks. Although principally applicable to all networks, systems thinking tools are especially relevant for self-driving networks that will rely on ML-based data-driven algorithms to autonomously drive networks---which can suffer from problems such as bias, noise, and unintended consequences---to help troubleshoot chronic problems and to ensure that no significant unintended consequences are ignored during design.

\bibliographystyle{ACM-Reference-Format}
\bibliography{paper}

\end{document}